\documentclass[preprint]{aastex}

\newcommand{\lsim}{\mbox{${{{}_{{}_<}^{~} \atop {}^{{}^\sim}_{~}}}$}}

\newcommand{\be}{\begin{equation}}
\newcommand{\ee}{\end{equation}}
\newcommand{\lab}[1]{\label{#1}}
\newcommand{\rr}[1]{~(\ref{#1})}
\def\gsim{{{}_{{}_>}^{~}\atop {}^{{}^\sim}_{~}}}

\begin{document}

\title{A Test for the Origin of Quasar Redshifts}

\author{Piotr Popowski and Wolfgang Weinzierl} 
\affil{Max-Planck-Institut f\"{u}r Astrophysik,
Karl-Schwarzschild-Str. 1, Postfach 1317, 85741 Garching bei
M\"{u}nchen, Germany} 
\affil{E-mails: popowski, weinzier@mpa-garching.mpg.de}

\begin{abstract} 
It is commonly accepted that quasar redshifts have a cosmological
character and that most of the quasars are at Gigaparsec
distances. However, there are some cases where several quasars with
completely different redshifts and a nearby active galaxy are aligned
in a certain way or occupy a very small patch on the sky, which is
claimed by some authors to be unlikely to happen by chance. 
Is there a small subset of quasars with non-cosmological redshifts?
For quasars apparently associated with galaxies, we consider
two scenarios for the origin of their redshift: 1. a change in the scale
factor of the whole Universe (standard, cosmological scenario), 
2. a velocity-induced Doppler shift of a nearby object's spectrum
(local, ejection scenario).
We argue for a simple astrometric test which can
distinguish between these two sources of quasar redshifts by
constraining their proper motions.

We give the predictions for the maximum possible proper motions of a
quasar for the cosmological and local scenarios of the origin of their
redshifts. We apply these theoretical results to the Bukhmastova
(2001) catalog, which contains more than 8000 close QSO-galaxy
associations.  In the standard interpretation of quasar redshifts,
their typical proper motions are a fraction of micro arc-second, and
beyond the reach of planned astrometric missions like GAIA and SIM. On the
other hand, the quasars ejected from local AGNs at velocities close to
the speed of light would have proper motions 5-6 orders of magnitude
larger, which would easily be measurable with future astrometric missions, or
even in some cases with HST, VLT and Keck telescope.  The distributions of
proper motions for the cosmological and local scenarios are very well
separated. Moreover, the division corresponds nicely to the expected
accuracy from GAIA and SIM.

Subject Headings: astrometry --- cosmology: miscellaneous ---
galaxies: active --- quasars: general --- relativity
\end{abstract}

\section{Introduction}

Quasi-stellar objects (QSO, quasars) were identified as a new class by 
Schmidt (1963). The first object with an understood spectrum was 3C
273, with a redshift of 0.158. 
Many more followed and very soon (Burbidge \& Burbidge 1967) it was 
obvious that there are no quasars with observed blueshifts. 
A wavelength shift in quasars can be attributed to:
\begin{enumerate}
\item a cosmological redshift, i.e. due to expansion of the whole
Universe (hereafter referred to as the cosmological scenario),
\item a gravitational redshift due to high concentration of material
in the emission region,
\item a pure Doppler shift associated with the rapid relative motions 
of the system quasar-observer (hereafter referred to as the local scenario),
\item some, as of yet undiscovered, phenomenon.
\end{enumerate}
Already in XIV century William Ockham realized that it is not necessary to
create new entities if the explanatory potential of the known entities is
sufficient --- therefore, we will exclude the last possibility from
the current discussion.
At the first glance, the absence of blueshifted quasars strongly
suggests that quasar redshifts have either a cosmological or
gravitational origin. The conditions needed to produce gravitational 
redshifts are so extreme that we do not discuss this possibility any further.
Historically, a cosmological character of redshifts was criticized
based on the argument that maintaining the observed quasar fluxes
requires huge energy supplies.
This issue is no longer perceived as a problem,
since the accretion of material on a central, super-massive black hole
provides a satisfactory explanation.
The Doppler (ejection) origin for quasars redshifts would be a viable 
alternative only if one could devise a way to avoid the blueshift
problem.
In the presence of only the longitudinal Doppler effect, about half of the
randomly-ejected objects should be blueshifted.
The situation, however, changes dramatically when the transverse Doppler
effect becomes important. In this case, many of the objects
approaching the observer are redshifted rather than blueshifted.
Still, in a significant population of quasars, one
should see a few objects that are blueshifted.

Current observational techniques have enabled well defined selection processes
for QSO's. Several large area sky surveys like 2 Degree Field and
Sloan Digital Sky Survey have
opened a new window in the identification of QSOs and allowed 
the acquisition of
a wealth of information (Croom et al.\ 2001; Schneider et al.\ 2002).
All of the thousands of quasars that we know of are redshifted.
It is unrealistic to expect that the ejection process would always
meet the conditions to produce only redshifts. Therefore, a universal
explanation of the redshifts of all quasars by the local 
hypothesis would require a giant conspiracy\footnote{For individual
quasars, the additional evidence for their cosmological redshift may 
come from the identification of their host galaxies in deep imaging or 
the detection of Lyman-$\alpha$ forest in their spectra. However, from
the ground, the Lyman-$\alpha$ detections are possible only for the 
quasars with $z \gsim 1.6$ (e.g., Kim et al.\ 2003), which is due to the 
cutoff associated with atmospheric absorption.}. But the ejection
mechanism may be responsible for the redshifts of a subset of QSOs.
There are some cases where several quasars with completely different 
redshifts and a nearby active galaxy occupy a very small patch on the sky, 
which is claimed to be unlikely to happen by chance (e.g., Arp 1999). 
Also, it is suggested that X-ray compact
sources tend to group around nearby active Seyfert galaxies (Burbidge,
Burbidge, \& Arp 2003). It is even
claimed by some that they preferentially lie along the minor axis 
of a Seyfert, and therefore were probably ejected from it 
(Chu et al.\ 1998, Arp 1999).
Here we do not engage in another discussion about the significance of
such cases.
However, we also do not follow the route of the overwhelming majority
of the astronomical community which ignores this evidence. Instead we
entertain the idea of a simple astrometric test which can produce
a clear, conclusive answer. We show that measuring the
proper motion ($\mu$) of QSOs or at least placing an upper limit on $\mu$
will allow one to distinguish between Doppler kinematic redshifts and the
cosmological ones.

Before presenting more detailed computations, it is instructive
to see why the measurements of quasar proper motions 
promise to distinguish between the local and cosmological
origin of quasar redshifts.
In the standard cold dark matter (CDM) cosmology the peculiar
velocities of high-redshift galaxies, are expected to be only
a tiny perturbation on top of the overwhelming Hubble flow.
The same is true for quasars because they form in the centers of
galaxies. Therefore, the peculiar velocities of quasars in the
cosmological scenario are likely to
be $v_{Q, \rm cosmo} \sim 10^{-3} c$, where $c$ is the speed
of light. On the other
hand, if the local ejections are to produce redshifts, the
ejection velocities must be close to the speed of light, i.e. $v_{Q,
\rm local} \sim c$.
In addition, the local quasars would be at a distance of
$D_{Q, \rm local} \sim 10$ Mpc, and the cosmological ones
at $D_{Q, \rm cosmo} \sim 1$ Gpc.
Consequently:
\be
\frac{\mu_{Q,\rm cosmo}}{\mu_{Q,\rm
local}}\sim(\frac{D_{Q, \rm cosmo}}{D_{Q,\rm local}})^{-1}(\frac{v_{Q,\rm
cosmo}}{v_{Q,\rm local}}) \sim 10^{-5} \lab{muRatio}   
\ee 
The idea of the astrometric test we present exploits this large
difference between the expected proper motions.
In \S 2 we briefly review the theory of relativistic ejection and derive
some new results. Future astrometric measurements that will allow the 
implementation of the proper motion test are discussed in \S 3.
In \S 4 we describe the catalog of QSO-galaxy associations that
we use, and in \S 5 derive the expected proper motions of quasars
in this catalog. We summarize the results in \S 6.  

\section{Theoretical considerations}

We start with a brief review of the kinematics of relativistic
ejection (Behr et al.\ 1976).
We note here that the mathematical formalism of relativistic ejection
can be used to describe both the local and the cosmological case.
In the local scenario, a quasar is ejected with a relativistic
velocity from an active galaxy.
In the cosmological scenario, a quasar may be viewed
as being ``ejected'' with its small peculiar velocity from the frame
of reference moving with the Hubble flow.
 
We introduce a cosmological frame of reference with the observer at
rest at its origin. The line connecting the observer and some point
$G$ (which may or may not represent a galaxy), defines the line of sight.
A quasar $Q$ is ejected from point $G$ with a velocity represented
by $\beta\equiv v/c$
as measured with respect to the local cosmological standard of rest
at $G$.
The angle of ejection $\alpha$ in the $G$ rest frame is the angle between 
the quasar velocity and the direction along the line of sight away
from the observer. 
With these definitions one obtains for the proper motion
$\mu$ of $Q$ with respect to $G$, 
\be \mu =
\frac{c}{D_M}\frac{\beta \sin\alpha}{1+\beta
\cos\alpha}.\label{functionmu} 
\ee 
where $D_M$ is the transverse comoving distance or proper motion distance.
At this point we choose standard $\Lambda$ cosmology with
$\Omega_M=0.3$ and $\Omega_{\Lambda}=0.7$ (e.g., Bennett et al.\
2003), and set the curvature contribution to the total mass
density $\Omega_K=0$. Then the transverse comoving distance 
is defined as 
\be D_M=D_H\int_0^z\frac{dz'}{E(z')},
\label{transdist}
\ee 
where the distance $D_H \equiv c/H_0$ is the Hubble
radius\footnote{Note that in equation\rr{transdist} the distance $D_M$ 
is expressed in terms of the redshift $z$ of point $G$, which we will
approximate by the measured redshift of a quasar in the
cosmological scenario or by the redshift of a galaxy in the local
scenario. This approximation leads to negligible errors in the 
cosmological scenario, but may be important for local galaxies,
since they have peculiar radial velocities that are often a substantial
fraction of the radial velocity of point $G$ with respect to the
observer. In some instances, when the appropriate observational data
exist, this problem may be remedied
by using measured distances to galaxies instead of the ones
derived from their redshifts.}, and we take the
Hubble constant to be $H_0 = h \times 100$ km/s/Mpc.
Under our assumptions, the function $E(z)$ (e.g., Hogg 2000) is given by:
\be 
E(z)=\sqrt{\Omega_M(1+z)^3+\Omega_{\Lambda}}.
\label{ez} 
\ee 

\begin{figure}[htb]
\includegraphics[width=16cm]{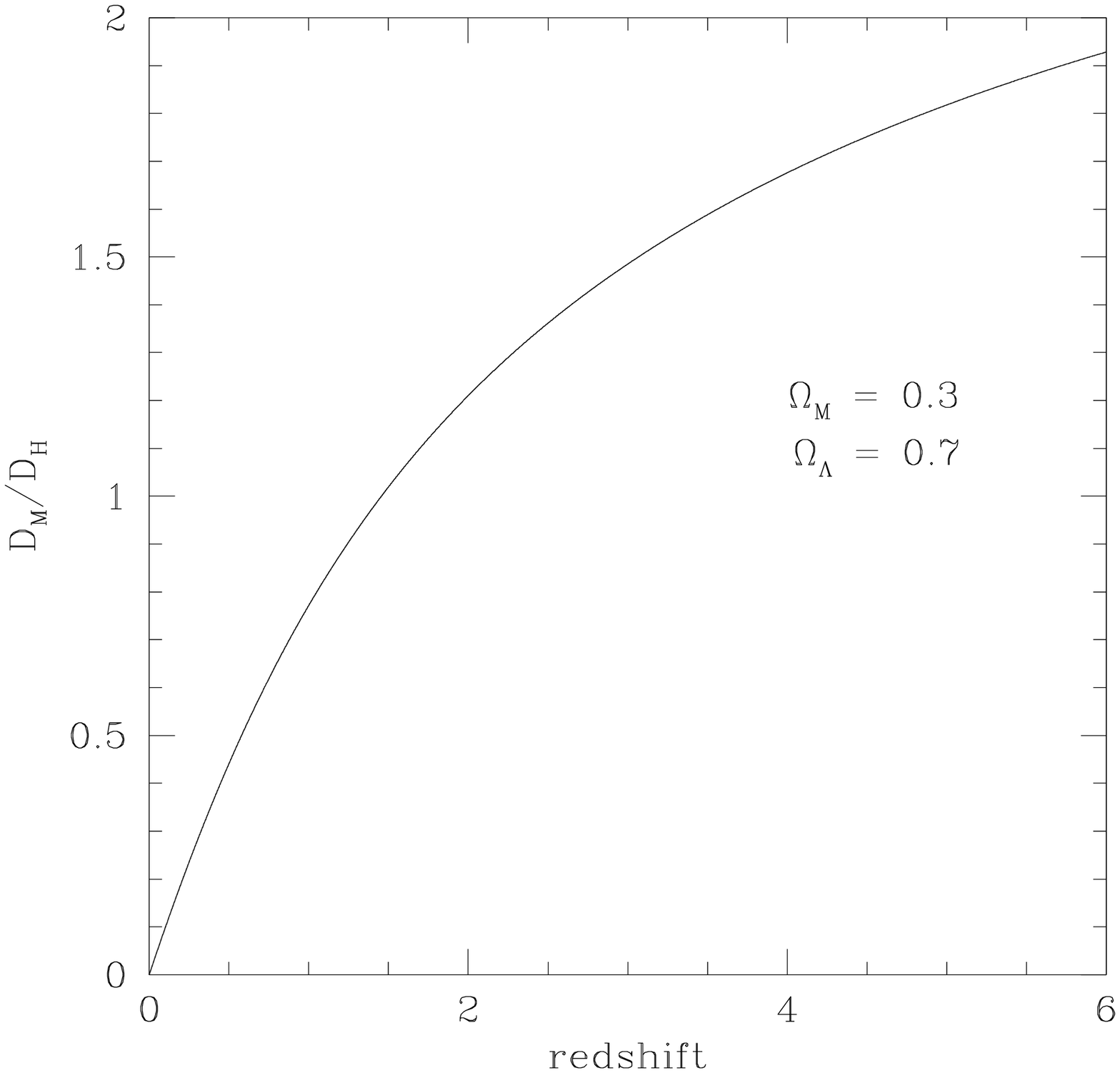}
\caption{The fraction $D_M/D_H$ as a function of redshift for $\Lambda$ 
cosmology with $\Omega_M=0.3$ and $\Omega_{\Lambda}=0.7$. \label{figure1}} 
\end{figure}

From\rr{functionmu} we see that the dependence of the proper motion on
the local geometry of ejection process is described by the function 
$F(\alpha,\beta)$.  
\be
F(\alpha,\beta)=\frac{\beta \sin\alpha}{1+\beta
\cos\alpha}.\label{ffunct}
\ee
The proper motion in the cosmological scenario can be derived from
Eq.\rr{functionmu}, if we associate ejection velocity with the
peculiar velocity of the quasar. In this case, we simply consider ``an
ejection'' from the reference frame moving with the Hubble flow.  For
a given $\beta=\beta_0$ a maximal proper motion is achieved for the
angle $\alpha_0$ such that $\cos\alpha_0=\beta_0$ and is equal to 
\be
\mu_{\rm max} = \frac{c}{D_M} \frac{\beta_0}{{(1-{\beta_0}^2)}^{1/2}}
\lab{mumaxcosmo} 
\ee 
We assume that the peculiar velocity does not
exceed $\beta_0=(1000 \, {\rm km/s})/c$.  This is very likely to be a
conservative assumption because the pairwise velocity dispersions at
$z \sim 0$ are of the order of a few hundred km/s (Jing \& B\"{o}rner 2001; 
Landy 2002) and they do not rise with $z$
for the currently favored $\Lambda$ cosmology (Barrow \& Saich 1993).
Hence, $\mu_{\rm cosmo}$ is very likely smaller than: 
\be 
\mu_{\rm cosmo, max} = 0.70367 \times 10^{-7} \times
\frac{D_H}{D_M} \; h^{-1} \, {\rm arcsec/yr}.\label{cosmopm}
\ee
The fraction $D_M/D_H$ as a function of redshift for $\Lambda$ cosmology with
$\Omega_M=0.3$ and $\Omega_{\Lambda}=0.7$ is displayed in Figure \ref{figure1}.

\placefigure{figure1}

In the local scenario the ejection is caused by a host galaxy. For an
ejected QSO (subscript $Q$) and the galaxy (subscript $G$), Behr et
al.\ (1976) define: 
\be
\kappa\equiv\frac{1+z_Q}{1+z_G}=\frac{1+\beta
\cos\alpha}{(1-{\beta}^2)^{\frac{1}{2}}}, \label{kappa}
\ee 
where $z_Q$ and $z_G$ are the redshifts of the QSO and galaxy,
respectively.  Since a typical quasar has $z_Q \sim 2$, and a typical
galaxy with which it is apparently associated has $z_G \ll 1$, the coefficient
$\kappa$ peaks around $3$.

\begin{figure}[htb]
\includegraphics[width=16cm]{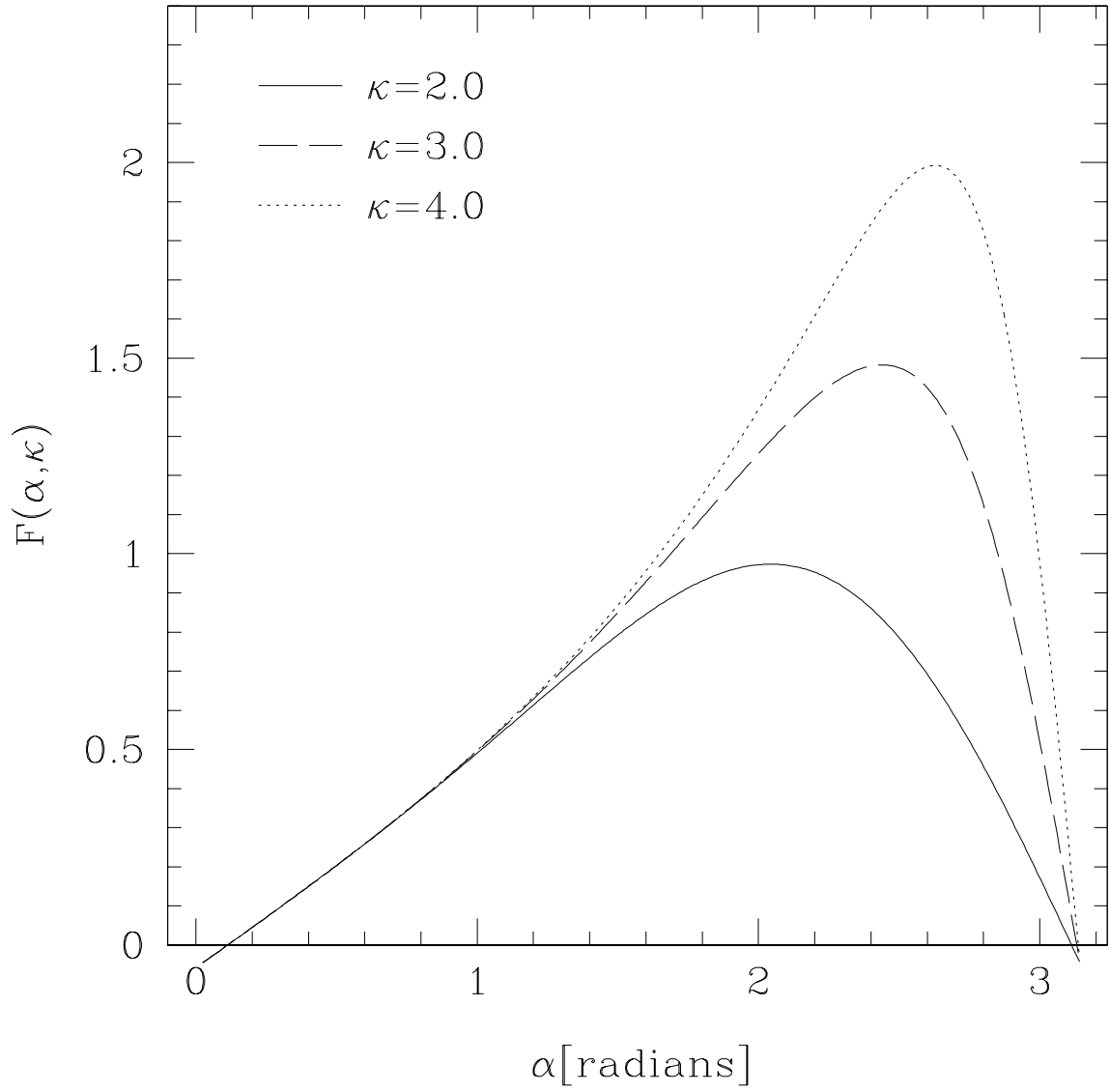}
\caption{The $\alpha$-dependence of $F(\alpha,\kappa)$ for
$\kappa = 2, 3, 4$.   \label{figure2}} 
\end{figure}

Our goal now is to derive the maximum possible proper motion given the
angle of ejection $\alpha$ and coefficient $\kappa$.  Eq.\rr{kappa}
yields 
\be
\beta=\frac{\cos\alpha+\kappa\sqrt{{\kappa}^2-\sin^2\alpha}}{{\kappa}^2+\cos^2\alpha}\label{beta}
\ee 
Substitution of eq.\rr{beta} into\rr{ffunct} yields
\be
F(\alpha,\kappa) = \frac{\sin\alpha
\cos\alpha+\kappa\sin\alpha
\sqrt{{\kappa}^2-\sin^2\kappa}}{{\kappa}^2+2\cos^2\alpha+\cos\alpha
\kappa \sqrt{{\kappa}^2-\sin^2\kappa}} \label{falphakappa} 
\ee 
We have plotted $F(\alpha,\kappa)$ as a function of $\alpha$ for
$\kappa = 2, 3$ and $4$ (Fig. \ref{figure2}). For each $\kappa$,
$F(\alpha,\kappa)$ peaks at a characteristic angle $\alpha_{\rm max}$
given by 
\be 
\alpha_{\rm max}= \left\{ \begin{array}{ll} {\rm
arccos}[\frac{\sqrt{4-4\kappa^2+\kappa^4}}{\sqrt{4+\kappa^2}}] & \mbox{if $ \kappa
\leq \sqrt{2}$,}\\ {\rm
arccos}[-\frac{\sqrt{4-4\kappa^2+\kappa^4}}{\sqrt{4+\kappa^2}}] & \mbox{if $ \kappa >
\sqrt{2}$.}  \end{array} \right.  \label{alphamax} 
\ee 
The algebraic manipulation of Eqs.\rr{falphakappa} and \rr{alphamax}
leads to the conclusion that the maximum observable proper motion in the
local scenario is given by: 
\be
\mu_{\rm local, max}=\frac{c}{D_M}\frac{\kappa}{2},\label{localpm} 
\ee
which is a very interesting new result.                  

\placefigure{figure2}
		
\section{Astrometric Missions}

The aim of this analysis is to design a test that will use the proper motions
to interpret quasar redshifts. Therefore, one should consider the most
powerful facilities that can make such a measurement.  We limit our
investigation to the upcoming astrometric missions 
GAIA\footnote{\tt http://astro.estec.esa.nl/GAIA/} and 
SIM\footnote{\tt http://planetquest.jpl.nasa.gov/SIM/sim\_index.html} that
are supposed to be launched around 2010.  Both will reach a parallax
(positional) measurement accuracy of about a few micro arc-seconds
($\mu$as) at visual magnitude $V=15$.  SIM will be able to point at
tens of thousands of
stars and GAIA will scan millions of stars over the entire sky.  Here
we describe the GAIA accuracy in some detail. We assume that SIM can
reach similar accuracy with the proper integration time; moreover we
expect that integration time that would allow SIM to surpass GAIA
substantially will in practice be very hard to obtain.  The parallax 
accuracy as a function of magnitude $m$ for a mission with a GAIA-like 
observational setup is given by Gould \& Salim (2002): 
\be 
\sigma_{\rm par} =
\sqrt{{\sigma_0}^2+{\sigma_{15}}^2\times10^{0.4(m-15)} \left[ 1+c_{RN}10^{0.4(m-15)}\right]},
\label{sigmapar}
\ee
where $\sigma_0=2.6 \, \mu$as, $\sigma_{15}=10.2\, \mu$as, and
$c_{RN}=0.012$ describe the
systematic-limited, photon-noise-limited, and readout-noise-limited
regimes for GAIA, respectively.
Gould \& Salim (2002) use $R$ as a representative magnitude. We apply
the same formula
to our data in $V$, and assume their parameters without any modification.

For $N$ observations with uniform sampling conducted over a duration of $t_E$,
the proper motion accuracy can be expressed in terms of $\sigma_{\rm
par}$ as 
\be 
\sigma_{\mu} =
2\sqrt{3}N^{-1/2}\sigma_{\rm par}/t_E.\label{sigmamu} 
\ee 
For the GAIA mission, it is predicted that on average $N = 82$ and
$t_E = 5$ years (Zwitter \& Munari 2003). The expected proper motion
errors obtained by combining equations\rr{sigmapar} and\rr{sigmamu}
are presented in Figure 3. The sampling will not be uniform for GAIA
mission but formula\rr{sigmamu} is a reasonable zeroth order approximation.

\begin{figure}[htb]
\includegraphics[width=16cm]{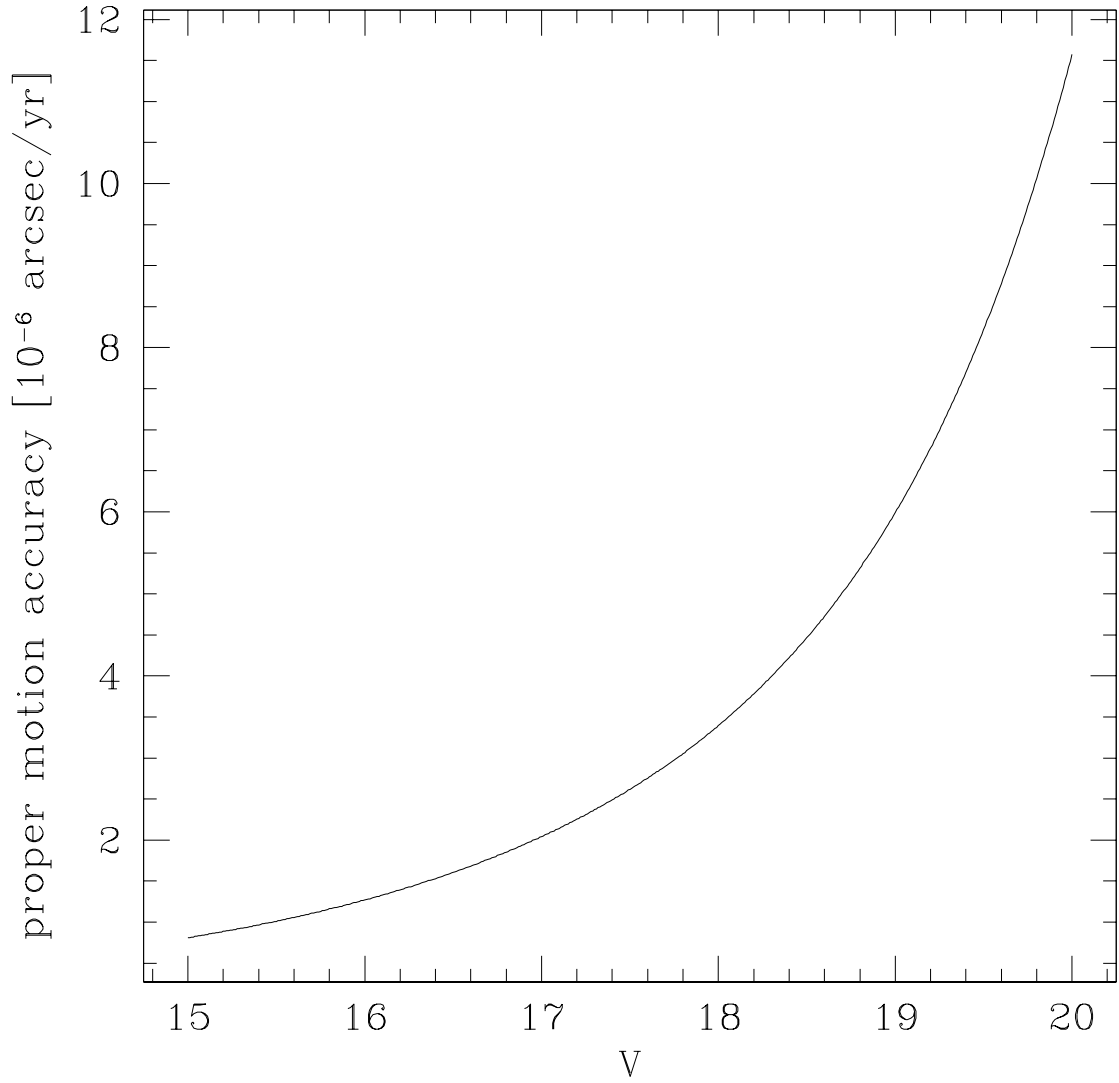}
\caption{GAIA proper motion errors under the assumption of uniform
sampling, the total number of observations equal to 82 and the duration of
the mission of 5 years. \label{figure3}}
\end{figure}

\placefigure{figure3}

\section{Data}

There have been a few quasar-galaxy associations discussed individually in
literature (Burbidge 1997, 1999; Burbidge \& Burbidge 1997; Chu et
al.\ 1998; Arp 1999). We have chosen a different approach and make a 
derivation of proper motions of a large number (8382) of close quasar-galaxy
pairs from the catalog by Bukhmastova (2001).  This catalog is based
on the Lyon-Meudon Extragalactic Database\footnote{\tt http://leda.univ-lyon1.fr/sample.html} that contains almost eighty
thousand galaxies
as well as the 8th edition of the Quasars and Active Galactic Nuclei
catalog by Veron-Cetty \& Veron (1998) that contains
11358 objects.  The associations were selected requesting:
(1) the availability of sky positions and redshifts for the quasars
and galaxies,
(2) the galaxy's redshift $z_G>0.0004$,
(3) redshift of QSO greater than the redshift of the
galaxy ($z_Q>z_G$), and
(4) the projected distance between the galaxy and QSO at the galaxy's
redshift $<150$ kpc
for $H_0=60$ km/s/Mpc.
The Bukhmastova (2001) catalog was our only source of data throughout 
the analysis.

\section{Results}

Using the theoretical results from \S 2, we have constructed a new
catalog which evaluates proper motions of quasars in QSO-galaxy 
associations. 
The complete version of this catalog, which will be published in the
electronic edition of the Journal, contains 8382 lines and is composed 
of 9 columns.  
We present 20 randomly chosen entries from this catalog 
in the printed version of Table 1. Columns 1 and 4 give the 
names of the QSO and galaxy, respectively, and columns 2 and 5 the 
redshifts of the QSO and galaxy, respectively. In column 3 we give 
the visual magnitude $V_Q$ of the QSO.  These were taken directly from the
catalog by Bukhmastova (2001). The next four columns contain the
quantities obtained by us.  In column 6 we list coefficients $\kappa$
[Eq.\rr{kappa}]. These are followed by the maximum proper motions for 
the local scenario $\mu_{\rm local, max}$ [Eq.\rr{localpm}] in column
7. We list $\mu_{\rm cosmo, max}$ [Eq.\rr{cosmopm}] in the 8th
column. Finally, the
expected proper motion accuracy for GAIA mission, $\sigma_{\mu}$
[Eq.\rr{sigmamu}], is presented in the last column.  This value depends
on the magnitude of a quasar, and should
be also representative of the achievable SIM error.  The expected GAIA
(achievable SIM) accuracy is given for 7126 entries in Table 1. The
remaining 1256 objects lack accuracy estimates due to one of the
following reasons: either $V_Q>20.0$ (1050 cases), which is fainter than
the formal limit of astrometric missions, or $V_Q$ was unknown (206
cases).

\placetable{table1}

\begin{figure}[htb]
\includegraphics[width=16cm]{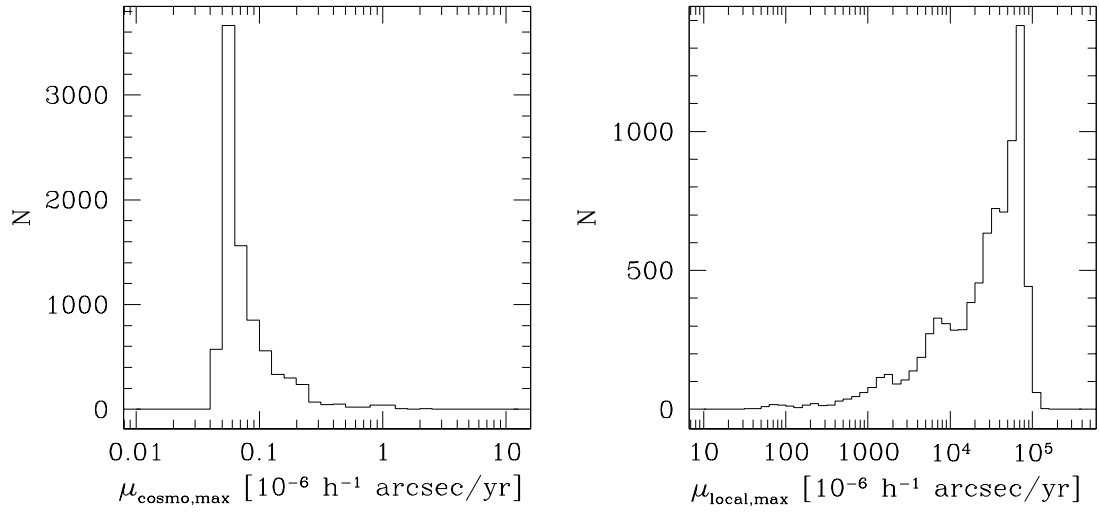}
\caption{The distribution of the maximum proper motions for the quasars
from the Bukhmastova (2001) catalog in two scenarios for
the origin of quasar redshifts.\label{figure4}} 
\end{figure}

\placefigure{figure4}

In Fig. \ref{figure4} we present the distribution of the number of quasars
with a given maximum proper motion. The left panel refers to the 
cosmological scenario and the right panel to the local ejection
scenario. Typical proper motion in the cosmological case are $\lsim
0.1$ $\mu$as/yr and the distribution of values is rather narrow.
This is a consequence of the fact that the number of quasars peaks
at $z \approx 2$. The distribution of expected proper motions
in the ejection scenario is wider with the proper motions expected
to be typically larger than a few milli arcsec/yr and reaching
0.1 arcsec/yr in the most extreme cases.

\begin{figure}[htb]
\includegraphics[width=16cm]{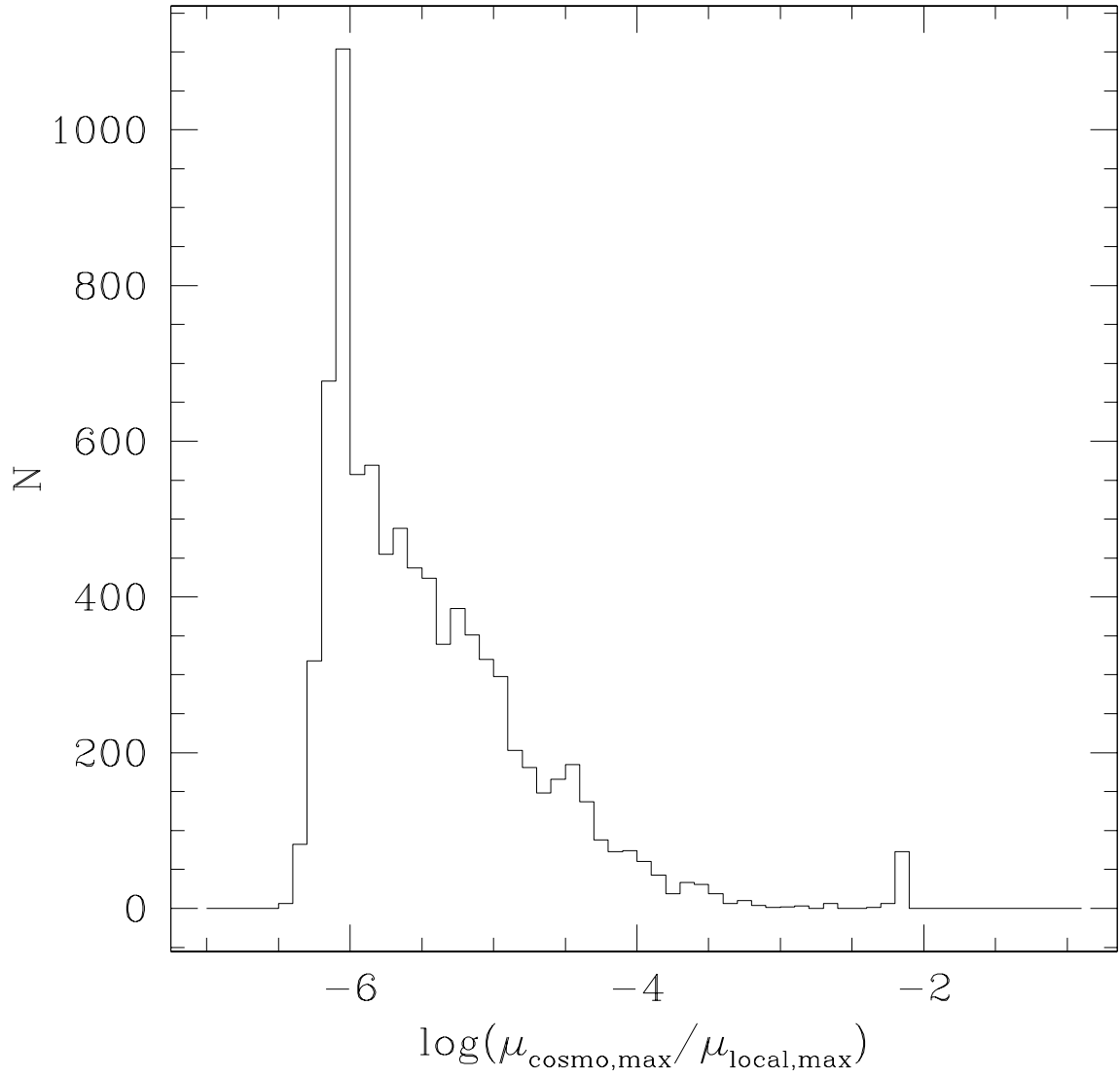} 
\caption{
The distribution of the ratio of quasar proper motions expected
from two scenarios for the origin of quasar redshifts.
\label{figure5}} 
\end{figure}

\placefigure{figure5}

Figure \ref{figure5} illustrates the ratio of the proper
motions in the cosmological and local scenarios on a
quasar-by-quasar basis. Typical quasars
have proper motions in the standard cosmological scenario that are
5-6 orders of magnitude smaller than those in the local scenario. 
This result confirms our simple expectation given in eq.\rr{muRatio}.
 
\begin{figure}[htb]
\includegraphics[width=16cm]{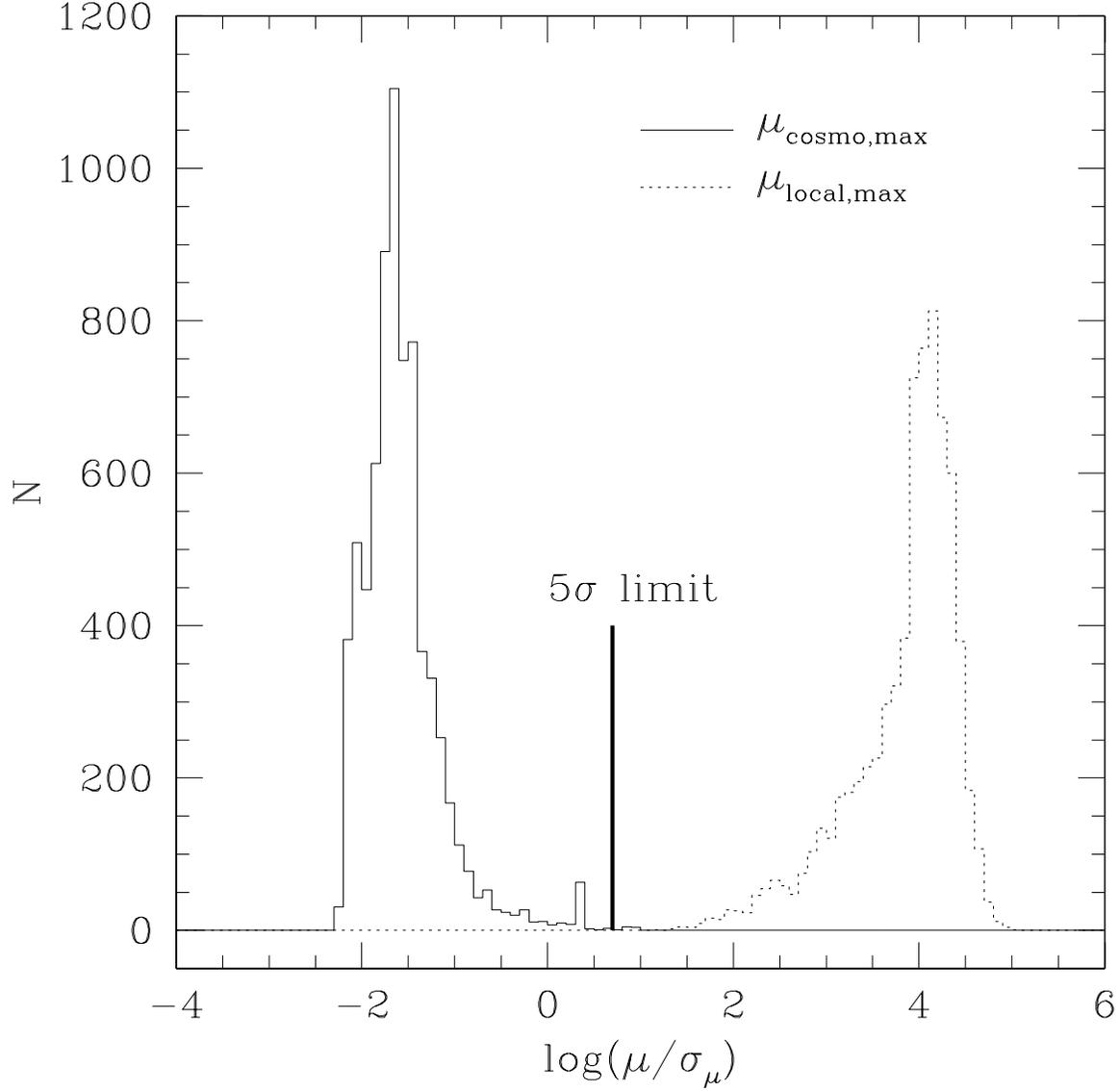}
\caption{The ratios of proper motions to the errors of
proper motion measurements expected from the GAIA astrometric
mission. For this plot, we assumed $h = 0.7$. The solid histogram
is based on the cosmological scenario and
the dotted one on the local ejection scenario. The GAIA 
$5\sigma$-detectability limit is marked with a vertical line.  
\label{figure6}} 
\end{figure}

\placefigure{figure6}

In Fig. \ref{figure6}, we plot the number of
associations as a function of the ratio of the proper motion and
astrometric errors. The solid histogram presents the results for the
cosmological interpretation of redshifts and the dotted histogram for the
local ejection scenario.  
We assume $H_0 = 70$ km/s/Mpc.
The vertical thick line marks the 5$\sigma$
detection limit.  For the local 
scenario, all 7126 maximum proper motions should be
detectable at the 5$\sigma$ limit.
Even if the proper motions are taken to be only one tenth of the
maximum value, GAIA should detect 99.7\% of them; if they are only
one hundredth of the maximum value, GAIA should still detect 94.4\%
of them.  
Figure \ref{figure2} shows that for a typical $\kappa \sim 3$, the proper
motion stays within a factor of a few from the maximum proper
motion for a wide range of angles $\alpha$.
Therefore, even if the objects have randomly
distributed ejection angles which would shift the histogram toward
lower signal-to-noise ratios, the proper motions should be detectable
in the great majority of instances.  
In the cosmological case, proper motions
are too small to be detected with present or presently
envisioned technology.
The proper motions for only 10 quasars from our sample (0.14\%) have a
chance to be detected at their cosmological distances. 
Nevertheless, the non-detections of proper
motions would undermine the kinematic interpretation of the
redshifts of quasars in QSO-galaxy associations,
giving support to the complete universality 
of the cosmological interpretation.

We warn that in some cases the apparent proper motion of
quasars may be associated with their variability rather 
than relativistic motion of an ejected object. Such
"false positives" could imply the local scenario even if 
the cosmological scenario is correct. Variability
of a quasar is the sum of the variations coming from the 
continuum-producing region, broad-line region, and 
narrow-line region. The typical sizes of these regions are 
$10^{-5}$ pc, 0.01 pc, and 10 pc, respectively (Peterson 1997). 
If we assume that the sources of the quasar's light can move by the 
characteristic sizes of the above mentioned regions in a coherent 
way, then the centroid of light from a given region can move by 
about $10^{-3}$, $1$, and $10^{3}$ $\mu$arcsec, respectively for 
a quasar at a distance of 1 Gpc. Therefore, any substantial 
contribution to flux variability from the narrow-line region 
may produce spurious detection of the quasar proper motion.

\section{Conclusion}

We have designed a proper motion test to distinguish between the
cosmological and Doppler (local ejection) origin of quasar
redshifts. We considered the process of relativistic ejection
and derived a new expression for the maximum proper motion
given the redshift of a galaxy $z_G$ and the redshift of
an ejected object (quasar) $z_Q$. 
This maximum proper motion depends only on the redshifts $z_G$ and $z_Q$
and the assumed cosmological model.

Using the Bukhmastova (2001) catalog of 8382 QSO-galaxy pair
associations we estimated typical proper motions for the two
considered scenarios. In the standard interpretation of quasar
redshifts, their typical proper motions are a fraction of micro
arc-second, and beyond the reach of planned astrometric missions like GAIA
and SIM.  On the other hand, the quasars ejected from local AGNs at
velocities close to the speed of light would have proper motions 5-6
orders of magnitude larger, and so easily measurable with future
astrometric missions, or even in some cases with HST, VLT and Keck
telescope\footnote{The measurements with HST, VLT and Keck
telescope would be probably possible only for the brightest quasars.
The HST Fine Guidance Sensor reaches a per-observation precision of $~ 1$ 
milli arcsec for $V < 16.8$ (\mbox{\tt http://www.stsci.edu/instruments/fgs}),
and NAOS/CONICA at VLT or NIRC at the Keck
telescope reach a few milli arcsec
precision from the diffraction limited photometry (e.g., Genzel et
al.\ 2003; Ghez et al.\ 2003).}. 
The distributions of proper motions for the cosmological and local
scenarios are very well separated. Moreover, the division corresponds
nicely to the expected accuracy from GAIA and SIM.

\acknowledgments
We thank Greg Rudnick for a very careful reading of the original
version of this manuscript and his helpful comments.

\clearpage

\clearpage

\begin{deluxetable}{lcclccccc} 
\tabletypesize{\scriptsize}
\tablecolumns{9} 
\tablewidth{0pc} 
\tablecaption{Proper Motions for Quasars in QSO-Galaxy Associations
\label{table1}} 
\tablehead{ 
\colhead{Quasar} & \colhead{$z_Q$} & \colhead{$V_Q$} &
\colhead{Galaxy} & \colhead{$z_G$} & \colhead{$\kappa$} &
\colhead{$\mu_{\rm local, max}$} & \colhead{$\mu_{\rm cosmo, max}$} &
\colhead{$\sigma_{\mu, \rm GAIA}$} \\ \colhead{} &\colhead{} &\colhead{}
&\colhead{} &\colhead{} &\colhead{} & \colhead{[$10^{-6} h^{-1} \rm{arcsec/yr}]$} &
\colhead{[$10^{-6} h^{-1} \rm{arcsec/yr}$]} & \colhead{[$10^{-6} \rm{arcsec/yr}$]}
}
\startdata   

MS 23574-3520      &$0.5080$ &$17.60$ &PGC0000621     &$0.0007$ &$1.507$ &$ 22707$  &$ 0.157$ &$    2.76$\\ 
PKS 2357-326       &$1.2750$ &$18.70$ &PGC0073049     &$0.0008$
&$2.273$ &$ 29971$  &$ 0.077$ &$    5.01$\\ 
TEX 2358+189       &$3.1000$ &$20.50$ &LEDA0138064    &$0.0024$ &$4.090$ &$ 18758$ &$ 0.047$ &too faint\\ 
Q 2359-397         &$2.0300$ &$19.00$ &PGC0001014     &$0.0004$ &$3.029$ &$ 79867$  &$ 0.058$ &$    5.99$\\ 
Q 0000-398         &$2.8270$ &$18.80$ &PGC0001014     &$0.0004$ &$3.825$ &$100876$\phm{1}  &$ 0.049$ &$    5.31$\\ 
Q 0000-4244        &$1.7000$ &$20.40$ &PGC0001014     &$0.0004$ &$2.699$ &$ 71169$ &$ 0.064$ &too faint\\ 
Q 0000-4239        &$2.1900$ &$21.10$ &PGC0001014     &$0.0004$ &$3.189$ &$ 84085$ &$ 0.055$ &too faint\\ 
Q 0001-4227        &$2.2400$ &$19.20$ &PGC0001014     &$0.0004$ &$3.239$ &$ 85403$  &$ 0.055$ &$    6.78$\\ 
Q 0001-4256        &$2.0300$ &$18.80$ &PGC0001014     &$0.0004$ &$3.029$ &$ 79867$  &$ 0.058$ &$    5.31$\\ 
Q 0001-4225        &$1.3000$ &$20.50$ &PGC0001014     &$0.0004$ &$2.299$ &$ 60625$ &$ 0.076$ &too faint\\ 
Q 0001-4255        &$2.0400$ &$18.40$ &PGC0001014     &$0.0004$ &$3.039$ &$ 80131$  &$ 0.058$ &$    4.22$\\ 
Q 0002-387         &$2.2300$ &$19.90$ &PGC0001014     &$0.0004$ &$3.229$ &$ 85139$  &$ 0.055$ &$   10.79$\phm{1}\\ 
Q 0002-422         &$2.7580$ &$17.21$ &PGC0001014     &$0.0004$ &$3.756$ &$ 99057$  &$ 0.049$ &$    2.27$\\ 
Q 0002-4305        &$2.2000$ &$19.80$ &PGC0001014     &$0.0004$ &$3.199$ &$ 84349$  &$ 0.055$ &$   10.06$\phm{1}\\
87GB 00250+4458    &$0.9710$ & \nodata &PGC0001777     &$0.0006$ &$1.970$ &$ 41554$ &$ 0.093$ &\nodata\\ 
Q 0025-4047        &$2.1800$ &$17.14$ &PGC0001014     &$0.0004$ &$3.179$ &$ 83821$  &$ 0.055$ &$    2.19$\\ 
Q 0025-4026        &$1.1730$ &$19.00$ &PGC0001014     &$0.0004$ &$2.172$ &$ 57278$  &$ 0.081$ &$    5.99$ 
\enddata
\tablecomments{The complete version of this table is in the electronic
edition of the Journal.  The printed edition contains only a sample.}
\end{deluxetable} 


\begin{references}
\reference{arp} Arp, H. 1999, \aap, 341, L5
\reference{bar} Barrow, J.D., \& Saich, P. 1993, \mnras, 262, 717 
\reference{beh} Behr, C., et al. 1976, \aj, 81, 3 
\reference{ben} Bennett, C.L., et al.\ 2003, \apjs, 148, 1
\reference{buk} Bukhmastova, Yu. L. 2001, \azh, 78, 675 
\reference{hog} Hogg, D. 2000, preprint (astro-ph/9905116 v4) 
\reference{bur1} Burbidge, E.M. 1997, \apj, 484, L99 
\reference{bur2} Burbidge, E.M. 1999, \apj, 511, L9
\reference{bur3} Burbidge, G., \& Burbidge, M. 1967, Quasi-stellar objects (W.H. Freeman and Company: San Francisco)
\reference{bur4} Burbidge, E.M., \& Burbidge, G. 1997, \apj, 477, L13
\reference{bur5} Burbidge, E.M., Burbidge, G., \& Arp, H. 2003, \aap,
400, L17
\reference{chu} Chu, Y., Wei, J., Hu, J., Zhu, X., \& Arp, H. 1998,
\apj, 500, 596
\reference{cro} Croom, S.M., Smith, R.J., Boyle, B.J., Shanks, T.,
 Loaring, N.S., Miller, L., Lewis, I.J. 2001, \mnras, 322, L29
\reference{gen} Genzel, R., et al.\ 2003, \apj, 594, 812
\reference{ghe} Ghez, A.M., Salim, S., Hornstein, S.D., Tanner, A.,
Morris, M., Beclin, E.E., \& Duche\^{e}ne, G. 2003, preprint (astro-ph/0306130)
\reference{gou} Gould, A., \& Salim, S. 2002, \apj, 572, 944
\reference{jin} Jing, Y.P., \& B\"{o}rner, G. 2001, \mnras, 325, 1389
\reference{kim} Kim, T.-S., Viel, M., Haehnelt, M.G., Carswell, R.F.,
\& Cristiani, S. 2003, preprint (astro-ph/0308103)
\reference{lan} Landy, S.D. 2002, \apj, 567, L1  
\reference{mun} Munari,U., \& Zwitter, T. 2003, preprint (astro-ph/0306019)
\reference{pet} Peterson, B.M. 1997, An introduction to active
galactic nuclei (Cambridge: University Press)
\reference{schm} Schmidt, M. 1963, Nature, 197, 1040
\reference{schn} Schneider, D.P., et al.\ 2002, \aj, 123, 567
\reference{ver} Veron-Cetty, M.-P., \& Veron, P. 1998, ESO Sci. Rep.,
18, 1 ({\tt \mbox{http://vizier.u-strasbg.fr/}
\mbox{viz-bin/VizieR?-source=VII/207}}) 
\end{references}
\end{document}